\documentclass[epj, nopacs]{svjour}
\usepackage{graphics}
\usepackage{hyperref}
\usepackage{amsmath,amssymb}
\newcommand{\dd}{\mbox{$\textrm{d}$}}
\newcommand{\pol}[1]{\mathaccent"017E{#1}}

\begin{document}
\hugehead
\markboth{P.~Adlarson et al., Backward single-pion production in the $pd\rightarrow {}^3\textrm{He}\,\pi^0$ reaction with
WASA-at-COSY}{P.~Adlarson et al., Backward single-pion production in the $pd\rightarrow {}^3\textrm{He}\,\pi^0$ reaction with
WASA-at-COSY}
\title{Backward single-pion production in the $\boldsymbol{pd\rightarrow {}^3\textrm{He}\,\pi^0}$ reaction with
WASA-at-COSY}
\titlerunning{Backward single-pion production in the $pd\rightarrow {}^3\textrm{He}\,\pi^0$ reaction with
WASA-at-COSY}

\author{{{The WASA-at-COSY Collaboration}\\[2ex]P.~Adlarson\inst{1}}\and
{W.~Augustyniak\inst{2}}\and
{W.~Bardan\inst{3}}\and
{M.~Bashkanov\inst{4}}\and
{F.S.~Bergmann\inst{5}}\and
{M.~Ber{\l}owski\inst{6}}\and
{A.~Bondar\inst{7,}\inst{8}}\and
{M.~B\"uscher\inst{9,}\inst{10}}\and
{H.~Cal\'{e}n\inst{1}}\and
{I.~Ciepa{\l}\inst{11}}\and
{H.~Clement\inst{12,}\inst{13}}\and
{E.~Czerwi{\'n}ski\inst{3}}\and
{K.~Demmich\inst{5}}\and
{R.~Engels\inst{14}}\and
{A.~Erven\inst{15}}\and
{W.~Erven\inst{15}}\and
{W.~Eyrich\inst{16}}\and
{P.~Fedorets\inst{14,}\inst{17}}\and
{K.~F\"ohl\inst{18}}\and
{K.~Fransson\inst{1}}\and
{F.~Goldenbaum\inst{14}}\and
{A.~Goswami\inst{14,}\inst{19}}\and
{K.~Grigoryev\inst{14,}\inst{20}}\and
{C.--O.~Gullstr\"om\inst{1}}\and
%{C.~Hanhart\inst{14,}\inst{21}}\and
{L.~Heijkenskj\"old\inst{1}\thanks{present address: Institut f\"ur Kernphysik, Johannes Gutenberg--Universit\"at Mainz, Johann--Joachim--Becher Weg~45, 55128 Mainz, Germany}}\and
{V.~Hejny\inst{14}}\and
{N.~H\"usken\inst{5}\thanks{Email: n.hues02@uni-muenster.de}}\and
{L.~Jarczyk\inst{3}}\and
{T.~Johansson\inst{1}}\and
{B.~Kamys\inst{3}}\and
{G.~Kemmerling\inst{15}\thanks{present address: J\"ulich Centre for Neutron Science JCNS, Forschungszentrum J\"ulich, 52425 J\"ulich, Germany}}\and
{G.~Khatri\inst{3}\thanks{present address: Department of Physics, Harvard University,
 17~Oxford St., Cambridge, MA~02138, USA}}\and
{A.~Khoukaz\inst{5}}\and
{A.~Khreptak\inst{3}}\and
{D.A.~Kirillov\inst{21}}\and
{S.~Kistryn\inst{3}}\and
{H.~Kleines\inst{15}\thanks{present address: J\"ulich Centre for Neutron Science JCNS, Forschungszentrum J\"ulich, 52425 J\"ulich, Germany}}\and
{B.~K{\l}os\inst{22}}\and
{W.~Krzemie{\'n}\inst{6}}\and
{P.~Kulessa\inst{11}}\and
{A.~Kup{\'s}{\'c}\inst{1,}\inst{6}}\and
{A.~Kuzmin\inst{7,}\inst{8}}\and
{K.~Lalwani\inst{23}}\and
{D.~Lersch\inst{14}}\and
{B.~Lorentz\inst{14}}\and
{A.~Magiera\inst{3}}\and
{R.~Maier\inst{14,}\inst{24}}\and
{P.~Marciniewski\inst{1}}\and
{B.~Maria{\'n}ski\inst{2}}\and
{H.--P.~Morsch\inst{2}}\and
{P.~Moskal\inst{3}}\and
{H.~Ohm\inst{14}}\and
{W.~Parol\inst{11}}\and
{E.~Perez del Rio\inst{12,}\inst{13}\thanks{present address: INFN, Laboratori Nazionali di Frascati, Via E.~Fermi, 40, 00044 Frascati (Roma), Italy}}\and
{N.M.~Piskunov\inst{21}}\and
{D.~Prasuhn\inst{14}}\and
{D.~Pszczel\inst{1,}\inst{6}}\and
{K.~Pysz\inst{11}}\and
{A.~Pyszniak\inst{1,}\inst{3}}\and
{J.~Ritman\inst{14,}\inst{24,}\inst{25}}\and
{A.~Roy\inst{19}}\and
{Z.~Rudy\inst{3}}\and
{O.~Rundel\inst{3}}\and
{S.~Sawant\inst{25}}\and
{S.~Schadmand\inst{14}}\and
{I.~Sch\"atti--Ozerianska\inst{3}}\and
{T.~Sefzick\inst{14}}\and
{V.~Serdyuk\inst{14}}\and
{B.~Shwartz\inst{7,}\inst{8}}\and
{K.~Sitterberg\inst{5}}\and
{T.~Skorodko\inst{12,}\inst{13,}\inst{27}}\and
{M.~Skurzok\inst{3}}\and
{J.~Smyrski\inst{3}}\and
{V.~Sopov\inst{17}}\and
{R.~Stassen\inst{14}}\and
{J.~Stepaniak\inst{6}}\and
{E.~Stephan\inst{22}}\and
{G.~Sterzenbach\inst{14}}\and
{H.~Stockhorst\inst{14}}\and
{H.~Str\"oher\inst{14,}\inst{24}}\and
{A.~Szczurek\inst{11}}\and
{A.~Trzci{\'n}ski\inst{2}}\and
{M.~Wolke\inst{1}}\and
{A.~Wro{\'n}ska\inst{3}}\and
{P.~W\"ustner\inst{15}}\and
{A.~Yamamoto\inst{28}}\and
{J.~Zabierowski\inst{39}}\and
{M.J.~Zieli{\'n}ski\inst{3}}\and
{J.~Z{\l}oma{\'n}czuk\inst{1}}\and
{P.~{\.Z}upra{\'n}ski\inst{2}}\and
{M.~{\.Z}urek\inst{14}}\and
{C.~Wilkin\inst{30}}}

\institute{
%1\address[IKPUU]
{Division of Nuclear Physics, Department of Physics and
 Astronomy, Uppsala University, Box 516, 75120 Uppsala, Sweden}\and
 %2\address[ASWarsN]
 {Department of Nuclear Physics, National Centre for Nuclear
 Research, ul.\ Hoza~69, 00-681, Warsaw, Poland}\and
%3\address[IPJ]
{Institute of Physics, Jagiellonian University, prof.\
 Stanis{\l}awa {\L}ojasiewicza~11, 30-348 Krak\'{o}w, Poland}\and
%4\address[Edinb]
{School of Physics and Astronomy, University of Edinburgh,
 James Clerk Maxwell Building, Peter Guthrie Tait Road, Edinburgh EH9 3FD,
 United Kingdom}\and
%5\address[MS]
{Institut f\"ur Kernphysik, Westf\"alische Wilhelms--Universit\"at
 M\"unster, Wilhelm--Klemm--Str.~9, 48149 M\"unster, Germany}\and
%6\address[ASWarsH]
{High Energy Physics Department, National Centre for Nuclear
 Research, ul.\ Hoza~69, 00-681, Warsaw, Poland}\and
%7\address[Budker]
{Budker Institute of Nuclear Physics of SB RAS, 11~akademika
 Lavrentieva prospect, Novosibirsk, 630090, Russia}\and
%8\address[Novosib]
{Novosibirsk State University, 2~Pirogova Str., Novosibirsk,
 630090, Russia}\and
%9\address[PGI]
{Peter Gr\"unberg Institut, PGI--6 Elektronische Eigenschaften,
 Forschungszentrum J\"ulich, 52425 J\"ulich, Germany}\and
%10\address[DUS]
{Institut f\"ur Laser-- und Plasmaphysik, Heinrich--Heine
 Universit\"at D\"usseldorf, Universit\"atsstr.~1, 40225 D\"usseldorf,
 Germany}\and
%11\address[IFJ]
{The Henryk Niewodnicza{\'n}ski Institute of Nuclear Physics,
 Polish Academy of Sciences, Radzikowskiego~152, 31--342 Krak\'{o}w,
 Poland}\and
%12\address[PITue]
{Physikalisches Institut, Eberhard--Karls--Universit\"at
 T\"ubingen, Auf der Morgenstelle~14, 72076 T\"ubingen, Germany}\and
%13\address[Kepler]
{Kepler Center f\"ur Astro-- und Teilchenphysik,
 Physikalisches Institut der Universit\"at T\"ubingen, Auf der
 Morgenstelle~14, 72076 T\"ubingen, Germany}\and
%14\address[IKPJ]
{Institut f\"ur Kernphysik, Forschungszentrum J\"ulich, 52425
 J\"ulich, Germany}\and
%15\address[ZELJ]
{Zentralinstitut f\"ur Engineering, Elektronik und Analytik,
 Forschungszentrum J\"ulich, 52425 J\"ulich, Germany}\and
%16\address[Erl]
{Physikalisches Institut, Friedrich--Alexander--Universit\"at
 Erlangen--N\"urnberg, Erwin--Rommel-Str.~1, 91058 Erlangen, Germany}\and
%17\address[ITEP]
%{Institute for Theoretical and Experimental Physics named
% by A.I.\ Alikhanov of National Research Centre ``Kurchatov Institute'',
% 25~Bolshaya Cheremushkinskaya, Moscow, 117218, Russia}\and
%17\address[ITEP]
{Institute for Theoretical and Experimental Physics named
 after A.I.\ Alikhanov of National Research Centre ``Kurchatov Institute'',
 25~Bolshaya Cheremushkinskaya, Moscow, 117218, Russia}\and
%18\address[Giess]
{II.\ Physikalisches Institut, Justus--Liebig--Universit\"at
 Gie{\ss}en, Heinrich--Buff--Ring~16, 35392 Giessen, Germany}\and
%19\address[IITI]
{Department of Physics, Indian Institute of Technology Indore,
 Khandwa Road, Simrol, Indore - 453552, Madhya Pradesh, India}\and
%20\address[HepGat]
%{High Energy Physics Division, Petersburg Nuclear Physics
% Institute, 188300 Gatchina, Russia}\and
%20\address[HepGat]
{High Energy Physics Division, Petersburg Nuclear Physics Institute
named after B.P.~Konstantinov of National Research Centre ``Kurchatov Institute'',
1 mkr.\ Orlova roshcha, Leningradskaya Oblast, Gatchina, 188300, Russia}\and
%21\address[IASJ]
%{Institute for Advanced Simulation, Forschungszentrum J\"ulich,
% 52425 J\"ulich, Germany}\and
%22\address[HeJINR]
{Veksler and Baldin Laboratory of High Energiy Physics,
 Joint Institute for Nuclear Physics, 6~Joliot--Curie, Dubna, 141980,
 Russia}\and
%23\address[Katow]
{August Che{\l}kowski Institute of Physics, University of
 Silesia, Uniwersytecka~4, 40--007, Katowice, Poland}\and
%24\address[NITJ]
{Department of Physics, Malaviya National Institute of
 Technology Jaipur, JLN Marg Jaipur - 302017, Rajasthan, India}\and
%25\address[JARA]
{JARA--FAME, J\"ulich Aachen Research Alliance, Forschungszentrum
 J\"ulich, 52425 J\"ulich, and RWTH Aachen, 52056 Aachen, Germany}\and
%25\address[Bochum]
{Institut f\"ur Experimentalphysik I, Ruhr--Universit\"at
 Bochum, Universit\"atsstr.~150, 44780 Bochum, Germany}\and
%26\address[IITB]
{Department of Physics, Indian Institute of Technology Bombay,
 Powai, Mumbai - 400076, Maharashtra, India}\and
%27\address[Tomsk]
{Department of Physics, Tomsk State University, 36~Lenina
 Avenue, Tomsk, 634050, Russia}\and
%28\address[KEK]
{High Energy Accelerator Research Organisation KEK, Tsukuba,
 Ibaraki 305--0801, Japan}\and
%29\address[ASLodz]
{Astrophysics Division, National Centre for Nuclear Research,
 Box~447, 90--950 {\L}\'{o}d\'{z}, Poland}\and
%30\address[Colin]
{Physics and Astronomy Department, UCL, Gower Street, London WC1E 6BT, United
Kingdom}}

\date{Received: date / Revised version: date}

%
\abstract{ New data on the production of single neutral pions in the
$pd\rightarrow{}^3\textrm{He}\,\pi^0$ reaction are presented. For fifteen proton
beam momenta between $p_p=1.60\;\textrm{GeV}/c$ and
$p_p=1.74\;\textrm{GeV}/c$, differential cross sections are determined over a
large fraction of the backward hemisphere. Since the only previous systematic
measurements of single-pion production at these energies were made in
collinear kinematics, the present work constitutes a significant extension of
the current knowledge on this reaction. Even this far above the production
threshold, significant changes are found in the behaviour of the angular
distributions over small intervals in beam momentum. \vspace{-1em}
%
%\PACS{{25.40.Qa} {$(p,\pi)$ reactions} \and {13.75.-n} {Hadron-induced low-
%and intermediate-energy reactions and scattering}}
} %end of abstract
\maketitle
%

%\linenumbers

\section{Introduction}
\label{intro}

\begin{figure}[htb]
	\resizebox{0.55\textwidth}{!}{
  		\includegraphics{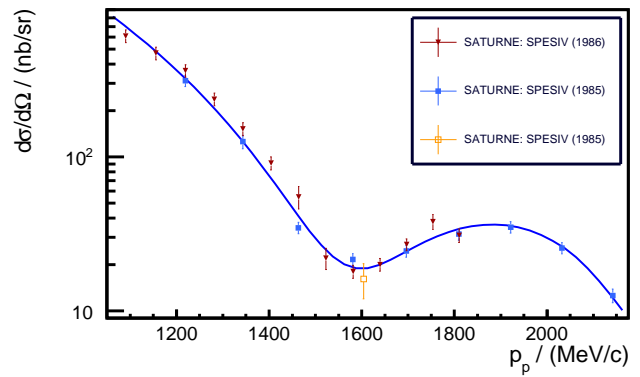}
	}
\caption{Differential cross sections of the reactions
$pd\rightarrow{}^3\textrm{He}\,\pi^0$ and
$pd\rightarrow{}^3\textrm{H}\,\pi^+$ (scaled by an isospin factor of $0.5$)
at $\cos\vartheta_{\pi^0}^*=-1$~\cite{Berthet:1985pw,Kerboul:1986qs}. The
curve represents a fourth order polynomial fit to the combined database.}
	\label{fig:pi0lit}
\end{figure}

In contrast to other meson production reactions in proton-deuteron fusion,
most notably the $\eta$- and $(\pi\pi)^0$-channels, the production of single,
neutral pions in the reaction $pd\rightarrow {}^3\textrm{He}\,\pi^0$ is
considerably less well-studied in the energy region around the
$\eta$-production threshold. Early measurements of the cross section (see
Fig.~\ref{fig:pi0lit}) and tensor analysing power of the $pd\;(dp)\rightarrow
{}^3\textrm{He}\,\pi^0$ and $pd\rightarrow {}^3\textrm{H}\,\pi^+$ reactions
in collinear kinematics with the SPES4 spectrometer at
SATURNE~\cite{Berthet:1985pw,Kerboul:1986qs} revealed strong structures in
both observables for backward pion production (with respect to the direction
of the incident proton in the c.m.\ frame) around $p_p=1.70\;\textrm{GeV}/c$.
It is important to note here that, since there is only one isospin amplitude,
the cross section for $pd\rightarrow {}^3\textrm{H}\,\pi^+$ should be twice
that for $pd\rightarrow {}^3\textrm{He}\,\pi^0$, though deviations of the
order of 10\% have been reported in the literature~\cite{LOW1981,ASL1977}.

Apart from a single measurement of $dp\rightarrow {}^3\textrm{He}\,\pi^0$ at
$p_d=3.5\;\textrm{GeV}/c$~\cite{Banaigs:1973ey}, angular distributions in the
$\eta$-threshold region have so far remained largely unexplored.
Nevertheless, the extensive database of cross sections for collinear
production, combined with the similarities of the $^3$He detection in the
reactions $pd\rightarrow {}^3\textrm{He}\,\eta$ and $pd\rightarrow
{}^3\textrm{He}\,\pi^0\pi^0$, have made $pd\rightarrow
{}^3\textrm{He}\,\pi^0$ a prime candidate for luminosity determinations in
fusion reactions~\cite{Adlarson:2014ysb,Adlarson:2014xmp}. There is, thus, a
twofold motivation for extended studies of differential cross sections of the
$pd\rightarrow {}^3\textrm{He}\,\pi^0$ reaction. These will permit an
exploration of the variations close to $\cos\vartheta_{\pi^0}^*=-1$ in the
vicinity of the $\eta$-production threshold as well as the establishment of a
new database for future experiments, that does not rely on an extrapolation
to collinear kinematics. Data obtained in parallel to the WASA-at-COSY
experiment on $\eta$-production away from threshold~\cite{Adlarson:2018rgs}
allow a detailed study of the cross sections for single-pion production over
a large part of the backward hemisphere.

\section{Experiment}
\label{sec:1}

The $\pi^0$ data were obtained at the WASA facility located within the Cooler
Synchrotron (COSY) of the Forschungs\-zentrum J\"ulich in the same experiment
as that designed to study $\eta$ production~\cite{Adlarson:2018rgs}. Beam
protons were steered to collide with pellets of frozen deuterium so that the
heavy ${}^3\textrm{He}$ ejectiles were emitted near the forward direction in
the laboratory frame. The WASA Forward Detector allows the energy and the
polar and azimuthal angles of the ${}^3\textrm{He}$ nuclei to be
reconstructed in multiple layers of plastic scintillators and a proportional
chamber, respectively. Detailed information on the experimental setup can be
found in Ref.~\cite{wasa}. Fifteen evenly spaced proton beam momenta were
used between $p_p=1.60\;\textrm{GeV}/c$ and $p_p=1.74\;\textrm{GeV}/c$, with
a resolution of $\Delta p/p\approx 10^{-3}$~\cite{Maier}. Utilising the
so-called supercycle mode of the accelerator, data can be taken at eight different
beam momentum settings, with multiple repeats one after another, thus minimising systematic
differences between the individual measurements. In practice, two such
supercycles were employed in this experiment. The measurement at
$p_p=1.70\;\textrm{GeV}/c$ was repeated in both supercycles and an additional
single-momentum measurement was made at $1.70\;\textrm{GeV}/c$ to allow
systematic effects between the two supercycles to be investigated.

\section{Analysis}
\label{sec:2}

\begin{figure}[ht!]
	\resizebox{0.45\textwidth}{!}{
  		\includegraphics{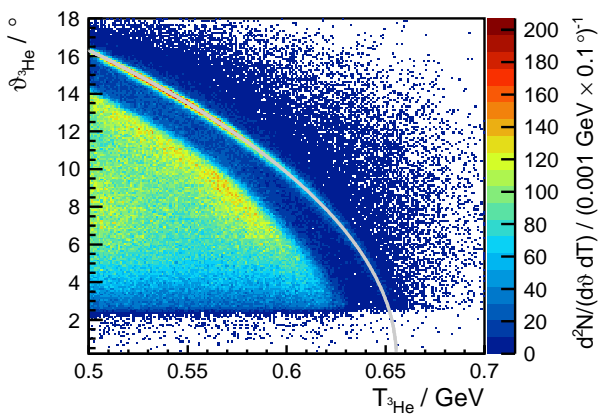}
	}
\caption{Production angle of ${}^3\textrm{He}$ nuclei in the laboratory frame displayed as a function
of its kinetic energy. The two-dimensional plot represents data (with color indicating the event yield) whereas the
grey line follows from applying four-momentum conservation to the
$pd\rightarrow {}^3\textrm{He}\,\pi^0$ reaction.}
	\label{fig:ekin}
\end{figure}

The ${}^3\textrm{He}$ nuclei produced near the forward direction are
identified in the WASA Forward Detector by means of their energy loss. From
this energy loss, a first value of the kinetic energy $T_{{}^3\textrm{He}}$
can be estimated by comparison with a Monte Carlo simulation of the
$\pi^0$ production reaction. By measuring also the polar and azimuthal
scattering angles $\vartheta$ and $\varphi$ in the Forward Proportional
Chamber, the four-momenta of the ${}^3\textrm{He}$ nuclei are fully
determined so that a missing-mass analysis can be performed. The analytic
relation between the precisely measured polar scattering angle
($\Delta\vartheta \approx 0.2^\circ$) and the kinetic energy of
${}^3\textrm{He}$ nuclei was exploited in order to carefully monitor the
energy calibration (see Fig.~\ref{fig:ekin}).

\begin{figure}[ht!]
	\resizebox{0.5\textwidth}{!}{
  		\includegraphics{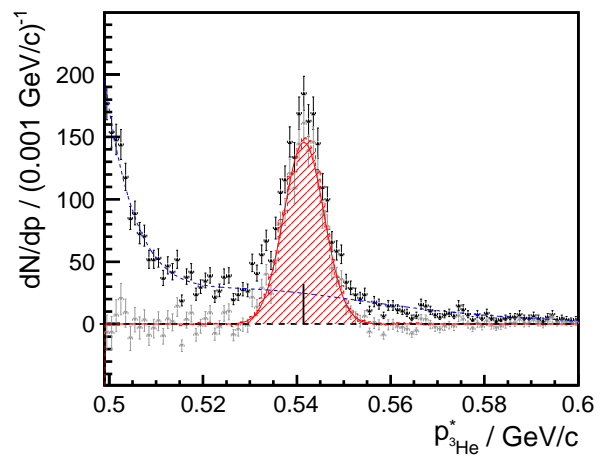}
	}
\caption{Spectrum of the final state momentum $p_{{}^3\textrm{He}}^*$ of
${}^3\textrm{He}$ nuclei in the centre-of-mass system for $-0.856\leq
\cos\vartheta_{\pi^0}^*< -0.840$ at a laboratory beam momentum of
$p_p=1.70\;\textrm{GeV}/c$. Black upward triangles represent data. The blue
dashed line is a fit of the type given in Eq.~(\ref{eq:fit}) to the spectrum,
excluding the peak region. Grey downward triangles show the data after
subtraction of the background fit. The $\pi^0$ peak is fitted by a Gaussian
distribution (continuous red line) and compared to a Monte Carlo simulation
of the $pd\rightarrow {}^3\textrm{He}\,\pi^0$ reaction (red shaded
histogram). The nominal peak position for a beam momentum of
$p_p=1.70\;\textrm{GeV}/c$ is indicated by the upright, solid black line.}
	\label{fig:fstm}
\end{figure}

For the study reported here, only the ${}^3\textrm{He}$ in the final state was used, 
even though the detector system was also capable of measuring photons from the decay of the $\pi^0$.
A missing-mass analysis allowed the angular distribution of single-pion
production to be derived from the final state momentum spectra binned in
$\cos\vartheta_{\pi^0}^*$. An example of such a final state momentum
spectrum, with a bin-width of $\Delta \cos\vartheta_{\pi^0}^* = 0.016$, can
be found in Fig.~\ref{fig:fstm}. The background, largely associated with two-pion
production and single-pion production with a poorly reconstructed energy due
to the breakup of ${}^3\textrm{He}$ nuclei in the scintillator material, is
subtracted using a fit of the type
\begin{equation}%
f(x)=e^{a(x-0.5)}(b+cx + dx^2), \label{eq:fit}%
\end{equation}
where $x$ is the $^3$He c.m.\ momentum. The fit is made outside the peak
region. The effect of nuclear breakup of the ${}^3\textrm{He}$ within the
detector is accounted for in a Monte Carlo simulation using an extension to
GEANT3~\cite{Geant3}, originally developed for the work in
Ref.~\cite{Bilger}. A fit of a Gaussian to the background-subtracted data is
used to define a $\pm3\sigma$ environment around the peak position. The event
yield within a certain bin in $\cos\vartheta_{\pi^0}^*$ is then defined as
the integral of the background-subtracted data in the $\pm3\sigma$ interval.

As no information is available on the angular distribution of single-pion
production, the product of acceptance and reconstruction efficiency (for
simplicity called below the \emph{acceptance} $A(\cos\vartheta_{\pi^0}^*)$)
is first derived from a Monte Carlo simulation of $\pi^0$ production,
assuming that the events are uniformly distributed over phase-space. The
angular distributions found in the experiment are corrected for the
acceptance by bin-wise multiplication with $A^{-1}(\cos\vartheta_{\pi^0}^*)$.
A polynomial fit of fourth order to these distributions is subsequently used
to weight the Monte Carlo simulations. This procedure is repeated until there
is convergence of $A(\cos\vartheta_{\pi^0}^*)$, when the angular
distributions are determined. This method was applied separately for the
measurements at all 15 beam momenta.

\begin{figure}[ht!]
	\resizebox{0.5\textwidth}{!}{
  		\includegraphics{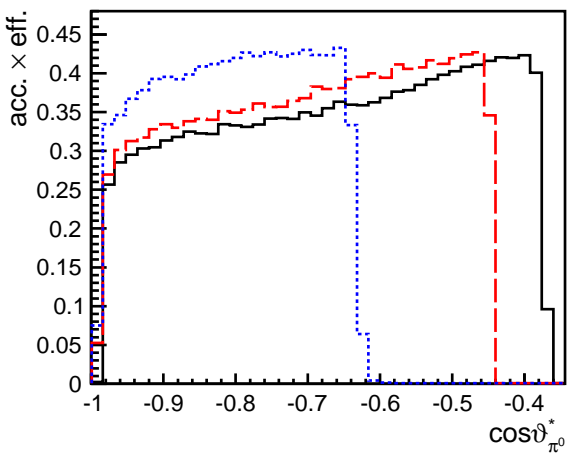}
	}
\caption{Product of acceptance and reconstruction efficiency
$A(\cos\vartheta_{\pi^0}^*)$ for the $pd\rightarrow {}^3\textrm{He}\,\pi^0$
reaction as function of the c.m.\ production angle for beam momenta of
$p_p=1.60\;\textrm{GeV}/c$ (blue, short-dashed line), $p_p=1.70\;\textrm{GeV}/c$ (red
long-dashed line) and $p_p=1.74\;\textrm{GeV}/c$ (black, solid line).}
	\label{fig:acc}
\end{figure}

Examples of the resulting acceptance as function of $\cos\vartheta_{\pi^0}^*$
are displayed in Fig.~\ref{fig:acc} for the measurements at
$1.60\;\textrm{GeV}/c$, $1.70\;\textrm{GeV}/c$ and $1.74\;\textrm{GeV}/c$. At
very large negative values of $\cos\vartheta_{\pi^0}^*$, the beam pipe in the
detector limits the acceptance, whereas for the smallest value of
$\vartheta_{\pi^0}^*$ the polar production angle $\vartheta_{{}^3\textrm{He}}$
exceeds the geometrical coverage of the Forward Detector. On average, the
angle-dependent acceptance $A(\cos\vartheta_{\pi^0}^*)$ is of the order of
40\%. This relatively small value is due mainly to the nuclear breakup of
the ${}^3\textrm{He}$ ions in the scintillator, leading to either a
misidentification or a poorly reconstructed kinetic energy.

\section{Normalisation}
\label{sec:3}

Since the data reported here were taken in parallel to those used to
investigate $\eta$ production in $pd$-fusion, the same normalisation
methodology can be applied~\cite{Adlarson:2018rgs}, and this is briefly
summarised. A first relative normalisation is derived from the ratio of
protons elastically scattered off the target deuteron at a certain beam
momentum and the reference momentum $p_p=1.70\;\textrm{GeV}/c$. The
luminosity for the measurement at $p_p=1.70\;\textrm{GeV}/c$ is obtained by
normalisation to the measured $pd\rightarrow {}^3\textrm{He}\,\eta$ total
cross section~\cite{Rausmann}. The statistical uncertainty of the luminosity
determination is of the order of $2\%$ but an additional systematic
normalisation uncertainty of $\approx 16\%$ needs to be considered. For more
information on the normalisation procedure, the reader is referred
to Refs.~\cite{Adlarson:2018rgs,mythesis}.

\section{Results}
\label{sec:4}

\begin{figure*}[ht!]
	\centering
	\resizebox{0.75\textwidth}{!}{
		\includegraphics{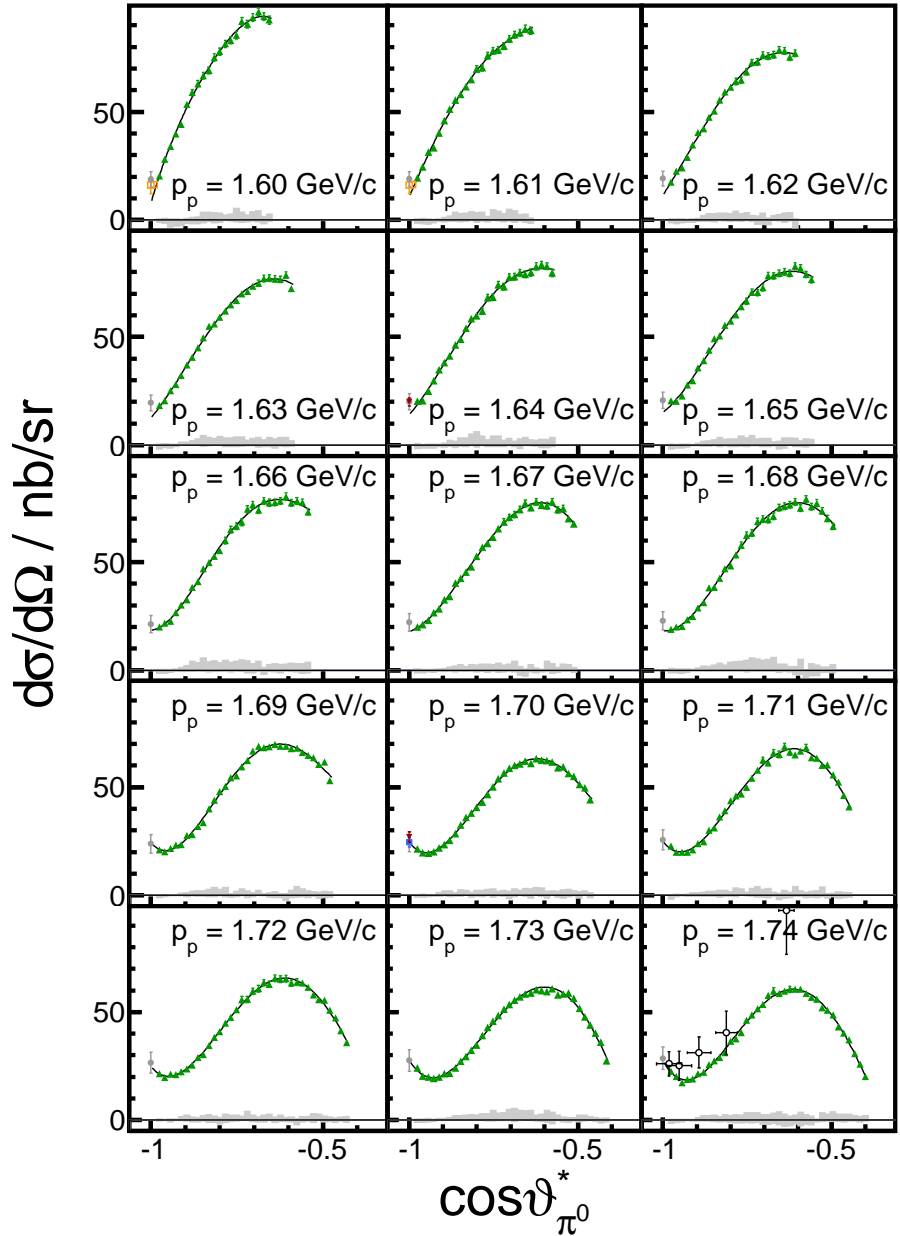}
	} \caption{Differential cross sections of the $pd\rightarrow
{}^3\textrm{He}\,\pi^0$ reaction in the backward hemisphere for the different
values of the beam momentum $p_p$ noted in each panel. Green upward triangles
represent the present data, yellow empty boxes a measurement at
$p_p=1.604\;\textrm{GeV}/c$~\cite{Berthet:1985pw}, the blue filled box a
measurement of $pd\rightarrow {}^3\textrm{H}\,\pi^+$ at
$p_p=1.697\;\textrm{GeV}/c$ (scaled by an isospin factor of
$0.5$)~\cite{Berthet:1985pw}, red downward triangles the measurements of
$dp\rightarrow {}^3\textrm{He}\,\pi^0$ at $p_d=3.276\;\textrm{GeV}/c$ and
$p_d=3.392\;\textrm{GeV}/c$~\cite{Kerboul:1986qs}, and the black empty
circles the measurement of $dp\rightarrow {}^3\textrm{He}\,\pi^0$ at
$p_d=3.50\;\textrm{GeV}/c$~\cite{Banaigs:1973ey}. In addition, the grey
filled circles represent the values of a fit to the combined database shown
in Fig.~\ref{fig:pi0lit} at the appropriate beam momentum. In the last case
the error bars are composed of the statistical uncertainty of the fit as well
as of the normalisation uncertainty of both the present experiment and the
literature data. Black solid lines represent fits of the type given in
Eq.~(\ref{fitsigma}). Normalisation uncertainties of the individual datasets are
not displayed. The grey histograms represent the systematic uncertainties,
other than normalisation, of the present measurements.}
	\label{fig:diffcross}
\end{figure*}

The angular distributions of single-pion production in proton-deuteron fusion
are converted into the differential cross sections shown in
Fig.~\ref{fig:diffcross} using the luminosities previously
derived~\cite{Adlarson:2018rgs,mythesis}. Apart from the normalisation, the
two main sources of the remaining systematic uncertainty, shown by the grey
histogram in Fig.~\ref{fig:diffcross}, are minor imprecisions in the
determination of the polar production angle (with a possible offset of
$\pm0.04^{\circ}$) and the distribution of residual gas within the WASA
scattering chamber. However, a comparison with the available data for collinear
production reveals a good agreement, especially if the normalisation
uncertainties of both the present data and those from the Saclay
experiments~\cite{Berthet:1985pw,Kerboul:1986qs} are taken into account. To
show this more clearly, the combined normalisation uncertainty is
incorporated into the error bars of the grey circles in
Fig.~\ref{fig:diffcross}.

The data shown in Fig.~\ref{fig:diffcross} can be well fit with the fourth
order polynomial
\begin{equation}
\label{fitsigma}
\frac{\dd\sigma}{\dd\Omega} = \sum_{n=0}^{n=4} a_n(\cos \vartheta_{\pi^0}^* + 1)^n
\end{equation}
and the values of the parameters $a_n$ are to be found in Table~\ref{tab:fitpar} for the
fifteen different beam momenta. Error bars are not shown because they are
very misleading in view of the very strong correlations between the parameters.
Nevertheless, the parametrisation of the data with Eq.~(\ref{fitsigma}) gives
a good description of our results that can be used in the normalisation of
other experiments.

For all the data above about 1.66~GeV/$c$ there is clear evidence for a
minimum in the differential cross section close to
$\cos\vartheta_{\pi^0}^*=-1$. However, the fit parameters of
Table~\ref{tab:fitpar} show that at lower momenta there may also be a minimum
but that it is in the unphysical region of $\cos\vartheta_{\pi^0}^*<-1$. The
minimum therefore moves somewhat with beam momentum from
$\cos\vartheta_{\pi^0}^*\approx -1.1$ at $p_p=1.62$~GeV/$c$ to $-0.93$ at
1.74~GeV/$c$. In contrast, the maximum seen in Fig.~\ref{fig:diffcross}
hardly moves from its position at $\cos\vartheta_{\pi^0}^*\approx -0.61$.

In general, six independent helicity amplitudes are required to describe
completely the $pd\rightarrow {}^3\textrm{He}\,\pi^0$ reaction. These reduce
to two, $A$ and $B$, in the forward and backward directions and the
magnitudes of these, $|A|^2$ and $|B|^2$, can be deduced from the Saclay
measurements of the differential cross section and deuteron tensor analysing
power $T_{20}$~\cite{Kerboul:1986qs}. At $\vartheta_{\pi^0}^*=180^{\circ}$ both
$|A|^2$ and $|B|^2$ show minima for proton beam momenta around 1600 to
1650~MeV/$c$. This behaviour, which causes rapid variations in $T_{20}$, must
clearly be linked to the moving minima seen in our data shown in
Fig.~\ref{fig:diffcross}.

\clearpage

\begin{table}[!ht]
	\centering
	\caption{Values of the parameters $a_n$ obtained from fits to our data
     using Eq.~(\ref{fitsigma}) for the different beam momenta.}
	\label{tab:fitpar}
	\begin{tabular}{ccrrrrr}%{cccccc}
	$p_p$  & $a_0$  & $a_1\phantom{1}$ & $a_2\phantom{1}$ & $a_3\phantom{1}$ & $a_4\phantom{1}$ & $\chi^2/$ndf \\
    GeV/$c$ & nb/sr&&&&&\\ \hline
	1.60 & \phantom{1}7.99 & 548 & -1566 & 4124 & -6096 & 14.0/16\\
	1.61 & 10.83 & 358 & -122 & -1447 & 1910 & 12.2/17\\
	1.62 & 11.63 & 231 & 523 & -2585 & 2117 & 16.4/19\\
	1.63 & 12.65 & 190 & 645 & -2464 & 1608 & 16.0/20\\
	1.64 & 14.25 & 171 & 793 & -2918 & 2294 & 19.9/21\\
	1.65 & 15.25 & 114 & 1024 & -3053 & 1968 & 28.0/22\\
	1.66 & 18.47 & 10 & 1787 & -5147 & 3886 & 17.3/23\\
	1.67 & 17.92 & 49 & 1067 & -2433 & 960 & 12.5/25\\
	1.68 & 18.38 & -30 & 1610 & -3776 & 2155 & 27.4/26\\
	1.69 & 24.43 & -188 & 2612 & -6461 & 4528 & 32.7/27\\
	1.70 & 24.43 & -200 & 2392 & -5601 & 3680 & 35.8/28\\
	1.71 & 25.04 & -203 & 2316 & -4978 & 2807 & 36.1/29\\
	1.72 & 24.67 & -188 & 2199 & -4779 & 2743 & 16.2/30\\
	1.73 & 27.70 & -254 & 2259 & -4434 & 2255 & 27.5/31\\
	1.74 & 27.47 & -285 & 2621 & -5545 & 3230 & 25.8/32\\ \hline
	\end{tabular}
\end{table}

It may also be relevant that the proton analysing power in the
$\pol{p}d\rightarrow {}^3\textrm{He}\,\pi^0$ reaction at fixed pion angle of
$\vartheta_{\pi^0}^* = 170^{\circ}$ shows an extremely rapid variation with the
proton beam momentum in the 1600~MeV/$c$ region~\cite{MAY1986}. There is
therefore much structure in large angle $pd\rightarrow
{}^3\textrm{He}\,\pi^0$ data near the $\eta$ threshold and it is tempting to
wonder whether this is more than an accident.

The amplitude for the $pd\rightarrow {}^3\textrm{He}\,\eta$ reaction near
threshold is anomalously strong~\cite{SMY2007,MER2007} and might be an
indication of the formation of a quasibound $_{\eta}^{3}$He
state~\cite{WIL1993}. There might therefore be an extra $S$-wave contribution
caused by $\eta$ production followed by the transmutation
$\eta{}^3\textrm{He}\rightarrow {}^3\textrm{He}\,\pi^0$ that will interfere
with a direct mechanism. However, using data on
$\pi^-{}^3\textrm{He}\rightarrow {}^3\textrm{H}\,\eta$~\cite{PEN1989}, it
seems that this two-step approach may be too small to explain the backward
minima. However, there is no valid reason to retain only the $^3$He ground
state in the intermediate state. Further theoretical work will be needed to
explore this interesting region.

\section{Summary}
\label{sec:5}

Measurements of the differential cross sections of single-pion production
in proton-deuteron fusion have been here reported for fifteen different
proton beam momenta between $1.60\;\textrm{GeV}/c$ and
$1.74\;\textrm{GeV}/c$. These data, which cover a large part of the backward
hemisphere, are a significant extension of the current database that
contained only detailed information for collinear production. Despite the
data being taken far above the $\pi^0$ production threshold, where the excess
energy is limited by $426\;\textrm{MeV} < Q_{\pi^0} < 494\;\textrm{MeV}$,
there are important changes in the angular distributions with
increasing $Q_{\pi^0}$. In particular at the lowest energy the large angle minimum is
at an unphysical point but it becomes observable with rising $Q_{\pi^0}$.
This and other phenomena~\cite{Kerboul:1986qs,MAY1986} seem to occur close to
the threshold for $\eta$ production. It will require much more theoretical
work to see if this is more than a coincidence.

Irrespective of the interpretation of the observed angular distributions, the
new data will be a valuable tool for normalising the cross sections for other
meson production reactions in proton-deuteron fusion. These data, which are
parametrised in Eq.~(\ref{fitsigma}) and Table~\ref{tab:fitpar}, will avoid having to rely
on any extrapolation to collinear kinematics.

\section*{Acknowledgements}

The work presented here received funding from the European Union Seventh
Framework Programme (FP7/2007-2013) under grant agreement number 283286. The
support given by the Forschungszentrum J\"ulich FFE Funding Programme of the
J\"ulich Centre for Hadron Physics, by the Polish National Science Centre
through grant No.\ 2016/23/B/ST2/00784, and by the DFG through the Research
Training Group GRK2149 is gratefully acknowledged. We thank the COSY crew for
their work and the excellent conditions made available during the beam time.

\clearpage

\begin{table*}[h!]
	\section*{Appendix}
	\centering
%	{\large \bf Appendix}
	\caption{Values in nb/sr of the $pd\to{}^3\textrm{He}\,\pi^0$ differential cross sections shown graphically in Fig.~\ref{fig:diffcross} from beam momenta of 1600~MeV/$c$ to 1670~MeV/$c$. The bins in 
$z=\cos\vartheta_{\pi^0}^*$ have width $0.016$ and the values quoted are the centres of these bins. Statistical uncertainties are given.\label{tab:results1} }
	\begin{tabular}{r|rrrrrrrr}
\hline
z\phantom{zz}&1600\phantom{zz}&1610\phantom{zz}&1620\phantom{zz}&1630\phantom{zz}&1640\phantom{zz}&1650\phantom{zz}&1660\phantom{zz}&1670\phantom{zz}\\
\hline
 -0.976&$20.4\pm1.0$&$19.5\pm1.0$&$17.6\pm0.9$&$18.3\pm0.9$&$20.3\pm1.0$&$20.4\pm1.0$&$20.0\pm1.1$&$20.1\pm1.0$\\
 -0.960&$28.1\pm1.2$&$24.7\pm1.0$&$22.6\pm1.0$&$20.5\pm0.9$&$20.5\pm1.0$&$20.3\pm0.9$&$21.7\pm1.0$&$21.2\pm0.9$\\
 -0.944&$34.1\pm1.2$&$31.3\pm1.1$&$24.1\pm1.0$&$25.2\pm1.0$&$24.8\pm1.0$&$22.7\pm0.9$&$22.7\pm1.0$&$23.2\pm1.0$\\
 -0.928&$39.9\pm1.3$&$33.3\pm1.1$&$29.1\pm1.1$&$27.9\pm1.0$&$29.8\pm1.1$&$27.8\pm1.0$&$26.7\pm1.1$&$26.6\pm1.0$\\
 -0.912&$44.2\pm1.3$&$40.3\pm1.1$&$34.7\pm1.1$&$32.1\pm1.0$&$34.9\pm1.1$&$29.7\pm1.0$&$30.2\pm1.1$&$28.3\pm1.0$\\
 -0.896&$53.4\pm1.4$&$45.9\pm1.2$&$40.5\pm1.2$&$37.0\pm1.1$&$38.1\pm1.2$&$35.7\pm1.1$&$32.6\pm1.2$&$32.5\pm1.1$\\
 -0.880&$58.8\pm1.5$&$51.2\pm1.3$&$42.1\pm1.2$&$40.6\pm1.1$&$41.3\pm1.2$&$39.1\pm1.1$&$38.3\pm1.2$&$34.2\pm1.1$\\
 -0.864&$63.0\pm1.5$&$55.5\pm1.3$&$47.4\pm1.3$&$45.1\pm1.2$&$46.2\pm1.3$&$43.9\pm1.2$&$41.0\pm1.3$&$40.4\pm1.2$\\
 -0.848&$66.6\pm1.5$&$58.1\pm1.3$&$50.3\pm1.3$&$49.7\pm1.2$&$48.9\pm1.3$&$49.1\pm1.3$&$46.9\pm1.3$&$42.6\pm1.2$\\
 -0.832&$69.2\pm1.6$&$61.6\pm1.4$&$55.4\pm1.3$&$54.9\pm1.3$&$54.1\pm1.4$&$50.6\pm1.3$&$49.8\pm1.4$&$45.3\pm1.3$\\
 -0.816&$75.1\pm1.6$&$64.9\pm1.4$&$59.1\pm1.4$&$56.0\pm1.3$&$58.3\pm1.4$&$55.4\pm1.3$&$52.6\pm1.4$&$47.9\pm1.3$\\
 -0.800&$77.9\pm1.7$&$69.8\pm1.5$&$61.3\pm1.4$&$59.1\pm1.3$&$59.7\pm1.4$&$57.4\pm1.4$&$55.3\pm1.4$&$52.6\pm1.3$\\
 -0.784&$81.4\pm1.7$&$70.4\pm1.5$&$64.0\pm1.5$&$62.0\pm1.4$&$61.9\pm1.4$&$60.3\pm1.4$&$60.0\pm1.5$&$56.8\pm1.3$\\
 -0.768&$83.0\pm1.7$&$75.9\pm1.5$&$64.7\pm1.4$&$64.7\pm1.4$&$68.1\pm1.5$&$64.0\pm1.4$&$65.0\pm1.5$&$58.6\pm1.4$\\
 -0.752&$85.4\pm1.7$&$78.2\pm1.5$&$68.5\pm1.5$&$66.8\pm1.4$&$68.3\pm1.5$&$66.5\pm1.5$&$66.6\pm1.5$&$61.4\pm1.4$\\
 -0.736&$91.7\pm1.8$&$78.6\pm1.5$&$72.8\pm1.5$&$70.0\pm1.4$&$74.1\pm1.5$&$70.7\pm1.5$&$68.4\pm1.6$&$65.4\pm1.4$\\
 -0.720&$90.5\pm1.8$&$80.3\pm1.5$&$73.1\pm1.5$&$71.1\pm1.4$&$73.2\pm1.5$&$70.5\pm1.5$&$74.4\pm1.6$&$68.5\pm1.4$\\
 -0.704&$93.6\pm1.8$&$83.6\pm1.6$&$76.2\pm1.5$&$73.4\pm1.4$&$77.6\pm1.6$&$72.7\pm1.5$&$76.4\pm1.6$&$70.7\pm1.5$\\
 -0.688&$96.2\pm1.8$&$85.6\pm1.6$&$75.6\pm1.5$&$74.8\pm1.4$&$77.9\pm1.6$&$78.3\pm1.5$&$74.0\pm1.6$&$72.0\pm1.5$\\
 -0.672&$93.9\pm1.8$&$86.5\pm1.6$&$76.4\pm1.5$&$77.0\pm1.5$&$79.6\pm1.6$&$78.2\pm1.5$&$78.2\pm1.6$&$74.4\pm1.5$\\
 -0.656&$92.4\pm1.8$&$88.3\pm1.6$&$78.5\pm1.5$&$77.2\pm1.5$&$79.0\pm1.6$&$79.4\pm1.5$&$77.4\pm1.6$&$77.0\pm1.5$\\
 -0.640&---\phantom{yyy}&$87.5\pm1.6$&$78.1\pm1.5$&$76.8\pm1.5$&$79.7\pm1.6$&$80.1\pm1.5$&$77.6\pm1.6$&$75.2\pm1.5$\\
 -0.624&---\phantom{yyy}&---\phantom{yyy}&$75.3\pm1.5$&$76.6\pm1.4$&$82.5\pm1.6$&$79.4\pm1.5$&$78.2\pm1.6$&$77.9\pm1.5$\\
 -0.608&---\phantom{yyy}&---\phantom{yyy}&$77.0\pm1.5$&$78.5\pm1.5$&$83.3\pm1.6$&$82.7\pm1.6$&$80.3\pm1.6$&$76.3\pm1.5$\\
 -0.592&---\phantom{yyy}&---\phantom{yyy}&---\phantom{yyy}&$72.4\pm1.4$&$82.7\pm1.6$&$81.8\pm1.5$&$77.4\pm1.6$&$76.0\pm1.5$\\
 -0.576&---\phantom{yyy}&---\phantom{yyy}&---\phantom{yyy}&---\phantom{yyy}&$79.5\pm1.6$&$78.9\pm1.5$&$78.2\pm1.6$&$77.8\pm1.5$\\
 -0.560&---\phantom{yyy}&---\phantom{yyy}&---\phantom{yyy}&---\phantom{yyy}&---\phantom{yyy}&$76.4\pm1.5$&$77.6\pm1.5$&$74.2\pm1.4$\\
 -0.544&---\phantom{yyy}&---\phantom{yyy}&---\phantom{yyy}&---\phantom{yyy}&---\phantom{yyy}&---\phantom{yyy}&$73.2\pm1.5$&$74.7\pm1.4$\\
 -0.528&---\phantom{yyy}&---\phantom{yyy}&---\phantom{yyy}&---\phantom{yyy}&---\phantom{yyy}&---\phantom{yyy}&---\phantom{yyy}&$70.2\pm1.4$\\
 -0.512&---\phantom{yyy}&---\phantom{yyy}&---\phantom{yyy}&---\phantom{yyy}&---\phantom{yyy}&---\phantom{yyy}&---\phantom{yyy}&$67.6\pm1.4$\\
\hline
	\end{tabular}
\end{table*}

\clearpage
\begin{table*}[!ht]
	\centering
		\caption{Values in nb/sr of the $pd\to{}^3\textrm{He}\,\pi^0$ differential cross sections shown graphically in Fig.~\ref{fig:diffcross} from beam momenta of 1680~MeV/$c$ to 1740~MeV/$c$. The bins in
$z=\cos\vartheta_{\pi^0}^*$ have width $0.016$ and the values quoted are the centres of these bins. Statistical uncertainties are given.\label{tab:results2} }
	\begin{tabular}{r|rrrrrrr}
\hline
z\phantom{zz}&1680\phantom{zz}&1690\phantom{zz}&1700\phantom{zz}&1710\phantom{zz}&1720\phantom{zz}&1730\phantom{zz}&1740\phantom{zz}\\
\hline
 -0.976&$19.0\pm1.0$&$21.0\pm1.0$&$21.6\pm0.6$&$22.8\pm1.0$&$21.4\pm1.2$&$24.1\pm1.1$&$22.6\pm1.1$\\
 -0.960&$19.9\pm1.0$&$20.2\pm0.9$&$19.8\pm0.5$&$19.9\pm1.0$&$19.7\pm1.1$&$20.3\pm0.9$&$20.2\pm1.0$\\
 -0.944&$20.2\pm1.0$&$21.7\pm0.9$&$19.5\pm0.5$&$20.0\pm1.0$&$21.3\pm1.0$&$19.9\pm0.9$&$17.2\pm0.9$\\
 -0.928&$23.4\pm1.0$&$23.2\pm0.9$&$20.1\pm0.5$&$20.0\pm0.9$&$21.1\pm1.0$&$19.3\pm0.9$&$18.8\pm0.9$\\
 -0.912&$24.8\pm1.0$&$23.5\pm0.9$&$21.9\pm0.5$&$21.0\pm0.9$&$22.3\pm1.0$&$19.9\pm0.9$&$19.0\pm1.0$\\
 -0.896&$28.9\pm1.1$&$27.7\pm1.0$&$23.6\pm0.6$&$23.9\pm1.0$&$23.7\pm1.0$&$21.5\pm0.9$&$21.5\pm1.0$\\
 -0.880&$31.2\pm1.1$&$28.2\pm1.0$&$25.5\pm0.6$&$24.8\pm1.0$&$25.3\pm1.1$&$21.7\pm0.9$&$22.1\pm1.0$\\
 -0.864&$38.1\pm1.2$&$31.9\pm1.1$&$28.5\pm0.6$&$30.1\pm1.1$&$29.0\pm1.1$&$24.0\pm1.0$&$25.6\pm1.0$\\
 -0.848&$38.4\pm1.2$&$33.6\pm1.1$&$32.4\pm0.6$&$31.1\pm1.1$&$30.3\pm1.1$&$27.3\pm1.0$&$27.4\pm1.1$\\
 -0.832&$42.4\pm1.3$&$39.9\pm1.2$&$36.1\pm0.7$&$36.3\pm1.2$&$34.5\pm1.2$&$29.3\pm1.1$&$29.1\pm1.1$\\
 -0.816&$46.9\pm1.3$&$44.1\pm1.2$&$37.9\pm0.7$&$39.2\pm1.2$&$38.2\pm1.3$&$32.2\pm1.1$&$31.3\pm1.1$\\
 -0.800&$48.2\pm1.3$&$47.9\pm1.2$&$40.8\pm0.7$&$43.8\pm1.3$&$40.9\pm1.3$&$34.6\pm1.1$&$36.4\pm1.2$\\
 -0.784&$53.4\pm1.4$&$50.3\pm1.3$&$43.2\pm0.7$&$46.1\pm1.3$&$44.9\pm1.3$&$39.0\pm1.2$&$37.6\pm1.2$\\
 -0.768&$57.7\pm1.4$&$54.3\pm1.3$&$46.6\pm0.7$&$46.8\pm1.3$&$47.5\pm1.4$&$42.9\pm1.2$&$40.5\pm1.2$\\
 -0.752&$62.3\pm1.5$&$55.3\pm1.3$&$50.2\pm0.8$&$50.0\pm1.3$&$51.0\pm1.4$&$45.7\pm1.2$&$46.3\pm1.3$\\
 -0.736&$63.4\pm1.5$&$59.4\pm1.4$&$53.8\pm0.8$&$54.0\pm1.4$&$55.7\pm1.5$&$48.5\pm1.3$&$48.6\pm1.3$\\
 -0.720&$68.8\pm1.5$&$62.3\pm1.4$&$56.5\pm0.8$&$57.9\pm1.4$&$55.9\pm1.4$&$51.3\pm1.3$&$51.6\pm1.4$\\
 -0.704&$69.8\pm1.5$&$66.6\pm1.4$&$58.8\pm0.8$&$62.4\pm1.4$&$59.6\pm1.5$&$53.6\pm1.3$&$53.4\pm1.4$\\
 -0.688&$69.5\pm1.5$&$68.8\pm1.4$&$60.0\pm0.8$&$61.1\pm1.4$&$60.8\pm1.5$&$55.4\pm1.3$&$57.8\pm1.4$\\
 -0.672&$71.1\pm1.5$&$68.2\pm1.4$&$60.8\pm0.8$&$65.8\pm1.5$&$63.6\pm1.5$&$57.0\pm1.3$&$59.8\pm1.4$\\
 -0.656&$75.2\pm1.6$&$68.7\pm1.4$&$62.0\pm0.8$&$65.2\pm1.5$&$62.9\pm1.5$&$58.4\pm1.3$&$58.6\pm1.4$\\
 -0.640&$75.7\pm1.6$&$69.8\pm1.4$&$60.9\pm0.8$&$68.7\pm1.5$&$65.6\pm1.5$&$58.9\pm1.3$&$59.7\pm1.4$\\
 -0.624&$76.5\pm1.6$&$68.9\pm1.4$&$63.2\pm0.8$&$66.2\pm1.4$&$65.2\pm1.5$&$60.4\pm1.4$&$60.5\pm1.4$\\
 -0.608&$77.7\pm1.6$&$68.8\pm1.4$&$62.5\pm0.8$&$64.9\pm1.4$&$65.4\pm1.5$&$59.9\pm1.3$&$60.5\pm1.4$\\
 -0.592&$74.9\pm1.5$&$67.7\pm1.4$&$62.2\pm0.8$&$66.7\pm1.4$&$63.4\pm1.5$&$59.5\pm1.3$&$60.5\pm1.4$\\
 -0.576&$79.1\pm1.6$&$68.1\pm1.4$&$61.5\pm0.8$&$68.3\pm1.4$&$63.9\pm1.5$&$60.7\pm1.3$&$58.6\pm1.4$\\
 -0.560&$75.6\pm1.6$&$65.9\pm1.3$&$59.1\pm0.8$&$63.6\pm1.4$&$63.5\pm1.4$&$57.8\pm1.3$&$56.8\pm1.4$\\
 -0.544&$77.2\pm1.6$&$64.6\pm1.3$&$59.4\pm0.8$&$63.4\pm1.4$&$61.3\pm1.4$&$58.4\pm1.3$&$56.0\pm1.3$\\
 -0.528&$73.7\pm1.5$&$63.5\pm1.3$&$56.4\pm0.7$&$60.0\pm1.3$&$58.0\pm1.3$&$58.8\pm1.3$&$52.0\pm1.3$\\
 -0.512&$70.1\pm1.5$&$60.4\pm1.2$&$54.9\pm0.7$&$60.4\pm1.3$&$55.8\pm1.3$&$56.3\pm1.3$&$50.9\pm1.2$\\
 -0.496&$66.4\pm1.4$&$61.5\pm1.2$&$51.1\pm0.7$&$55.6\pm1.3$&$55.6\pm1.3$&$51.7\pm1.2$&$48.5\pm1.2$\\
 -0.480&---\phantom{yyy}&$53.0\pm1.2$&$49.7\pm0.7$&$52.4\pm1.2$&$50.7\pm1.2$&$47.7\pm1.1$&$43.1\pm1.1$\\
 -0.464&---\phantom{yyy}&---\phantom{yyy}&$44.2\pm0.6$&$46.2\pm1.1$&$47.3\pm1.2$&$43.9\pm1.1$&$40.2\pm1.1$\\
 -0.448&---\phantom{yyy}&---\phantom{yyy}&---\phantom{yyy}&$40.8\pm1.0$&$41.4\pm1.1$&$40.3\pm1.1$&$37.2\pm1.0$\\
 -0.432&---\phantom{yyy}&---\phantom{yyy}&---\phantom{yyy}&---\phantom{yyy}&$35.8\pm1.0$&$36.0\pm1.0$&$30.6\pm0.9$\\
 -0.416&---\phantom{yyy}&---\phantom{yyy}&---\phantom{yyy}&---\phantom{yyy}&---\phantom{yyy}&$27.4\pm0.9$&$25.9\pm0.8$\\
\hline
	\end{tabular}
\end{table*}

\end{document}